# Indications for extra scalars at LHC ?
# BSM physics at future e+e- colliders

François Richard[1]

Université Paris-Saclay, CNRS/IN2P3, IJCLab[2], 91405 Orsay, France

______________________________________________________________

**Abstract:** *This paper intends to collect available data on searches for scalar resonances at LHC. It is suggested that, in the absence of SUSY, the most compelling picture is the composite framework, with the idea that the lightest particles are composite scalars of the pseudo-Nambu-Goldstone type, emerging from a broken symmetry at a higher scale, the h(125) boson being one of them. Searches in two-photons, Z-photon, ZZ into 4 leptons, top, h and W pairs are reviewed. A recent search based on lepton tagging from a spectator W/Z is also discussed. Aside from the already well-known scalar observed by CMS and LEP2 at 96 GeV, I present the evidence and the interpretation for a possible resonance observed in ZZ around 700 GeV by CMS and ATLAS and for a CP-odd scalar at ~400 GeV.*
*Future searches at HL-LHC and at e+e- colliders are briefly sketched.*

## I.    Introduction

After the Higgs boson discovery and the SUSY non-discovery, there is no guiding principle for future searches in HEP, no compelling direct evidence for new resonances or new phenomena guiding our choice for the energy needed at future colliders. Some hints exist, individually with low significance but, perhaps, put together indicating a coherent picture. The purpose of this note is to focus on scalar resonances, to see how they could be interpreted, within compositeness as pseudo-Nambu-Goldstone bosons, **pNGB**, and how they could justify our need for future colliders, with particular emphasis for e+e- projects.

Giving up on SUSY, one can take various attitudes:

- Avoid any specific theory and describe the measurements within an '*Effective Field Theory*' approach, EFT, which provides a sophisticated parametrization of the unknown new interactions. Whether this approach alone will lead to precise predictions on BSM particles, a requisite to design future colliders, remains an open question

---
[1] richard@lal.in2p3.fr
[2] Laboratoire de Physique des 2 Infinis Irène Joliot-Curie

- Assume that h(125) is **composite**, in very much the same way as pions are the lightest composite particles in QCD
- Assume that there is an **extra warped dimension** which allows to generate the hierarchy between the Planck mass and the EW scale
- Within extra dimensions, consider the Higgs boson as an extra component of the standard gauge particles, the so called **Gauge-Higgs unification** picture which has the virtue of protecting the Higgs sector from arbitrariness and large quartic divergences

The last three mechanisms are presumably deeply connected.

Without going into technical details, let me try to sketch an emerging general picture which could encompass these various ideas and allow some guidance for direct searches at LHC.

## I.1 Light pseudo-Nambu-Goldstone bosons

If one assumes that the Higgs boson is a composite particle, one gets rid of the so-called hierarchy problem induced by the nasty quartic mass divergence of an elementary scalar assumed in the SM. In this composite model, one also expects new heavy resonances, the analogue of the QCD vector resonances, so far not observed at LHC.

To explain why only the Higgs boson is light, one assumes that it is a pseudo **Nambu-Goldstone boson, pNGB**, emerging from a broken symmetry at a higher scale, in analogy to the usual pion observed in QCD. As in QCD, one can expect several boson like $\pi$, $\eta$ … This symmetry is of course unknown, which allows freedom in the predictions and therefore many open possibilities to generate extra light scalars.

In detail, note that the proponents of this model can also explain the pattern of couplings of h(125) which is totally similar to the SM Higgs boson.

In this picture, the discovery of a light Higgs boson at LHC appears very promising since it indicates the presence of a new layer of interactions at a higher unknown mass scale, which we can already reach thanks to the symmetry breaking mechanism.

It is thus natural to ask whether there are **other light scalars than h(125).** As within SUSY, these scalars could be **iso-singlets** or organized in **iso-doublets models** to insure the fulfilment of precision constraints. The partons constituting the scalar pNGB could have different quantum numbers like, for instance for scalar leptoquarks which, in some models, are held responsible for the anomalies observed in the B sector.

On the cosmological front, one knows that the SM framework alone cannot provide the **first order EW transition** needed to understand our universe. Having several scalars may provide the necessary ingredients to achieve this feature.

At LHC, the cleanest search channels for these scalars are the 4 leptons from a decay into ZZ, the two-photons and the Z-photon final states. Less clean and with poor mass resolution, but of potential interest, are top and W pairs in leptonic final states. In the following, I will go through the present evidences for scalar particles.

Before starting this analysis, a warning is of order. Although I am using the **familiar vocabulary** inherited from the usual MSSM, I would like to stress that my approach can radically differ from this framework. I am searching for **composite resonances**, which could be bound state of BSM partons or even result from bound states from standard bosons like ZZ or hh, in which case the decay patterns can **radically differ** from what is familiar with the MSSM particles.

## II. X(96)->bb at LEP2 and in two-photons at CMS

Two effects were observed at LEP2 and CMS at a mass of 96 GeV. The effect at LEP2 comes from a search for the HZ mode in a four-jets analysis combining the 4 experiments [1]. The evidence is below 3 standard deviation (sd). At CMS the analysis in two-photons revealed an excess in the two-photon mode at the same mass [2]. There are multiple interpretations of this effect, either in terms of NMSSM, where this resonance originates from a **Higgs singlet** or in terms of a **radion,** which appears in the Randall Sundrum scheme [3]. The measured width of this resonance is within the experimental resolution.

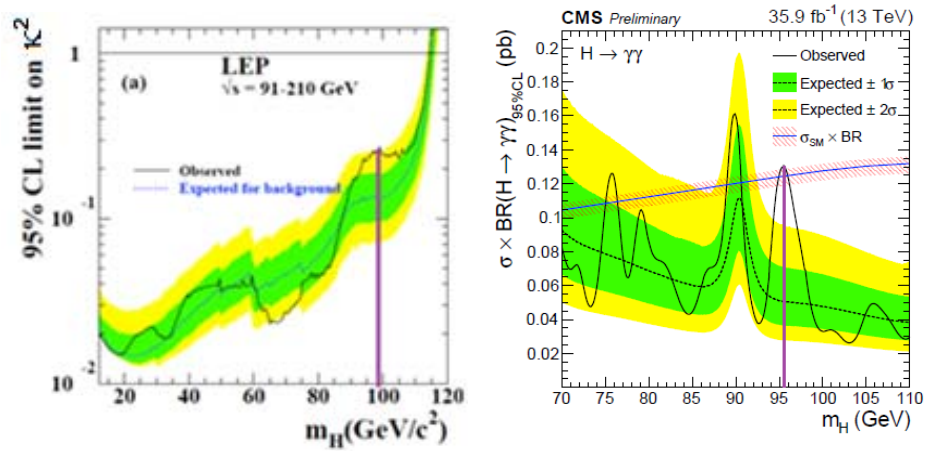

Figure 1: 95% CL limits from LEP2 and CMS on Higgs-like scalars versus mass. For CMS, the hatched curve corresponds to a SM Higgs coupling. For LEP2, $\kappa^2=1$ corresponds to a SM Higgs coupling.

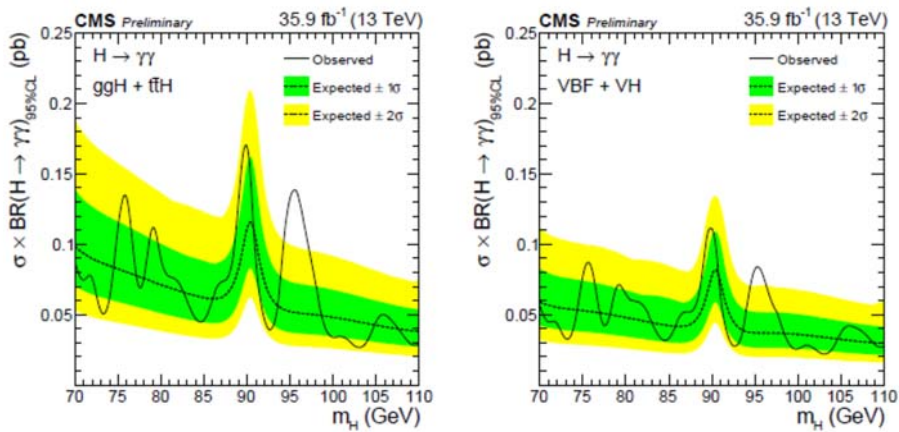

Figure 2: Comparison of 95% CL limits for ggF and VBF on Higgs-like scalars versus mass.

VBF (Vector Boson Fusion) indication from figure 2 suggests that this particle couples to ZZ/WW. For a SM Higgs boson, one expects a factor 10 between ggF and VBF [4], while one observes a reduction of only a factor 2, with a large error. This constitutes a puzzle w/o any obvious solution:

- Increasing VBF by increasing the X(96)ZZ coupling would require decreasing BR(bb) to explain the LEP2 result which is far from obvious since bb is expected to be the dominant mode. A

- possibility would be to assume that BR(bb) is reduced, due to extra decay modes, e.g. invisible, with large contributions
- Decreasing $\Gamma_{gg}$, which would request increasing BR($\gamma\gamma$) to achieve the correct ggF cross-section for LHC, seems more practical if X(96) behaves as a composite particle rather than as a genuine Higgs (see the interpretation section IV)

By how much should we increase BR ($\gamma\gamma$) (and decrease $\Gamma_{gg}$ accordingly), to agree with CMS? We measure a ratio 2 instead of the predicted 10. Recall also that if we do not modify BR(bb), we need to compensate for WW/ZZ fusion to take into account the $\kappa^2$ reduction factor measured at LEP2 (figure 1). This would mean another factor ~5, hence ~25 in total.

For $\Gamma_{gg}$, which is proportional to the square of the Yukawa coupling, this would mean dividing this coupling Yukawa $Y_t$ by 5, to a value $Y_t$~0.2 instead of 1. For BR($\gamma\gamma$) a factor 25 increase would mean to bring the BR from 0.15% to 3.7%, which does not seem to be a problem in a composite model.

It is improbable that LHC will measure X(96) into bb, since it is almost mass degenerate with Z. LHC will therefore be unable to understand separately the two factors governing the LHC rate, $\Gamma_{gg}$ and BR(bb). By measuring the fusion process, it will provide some information on $\Gamma_{WW}$.

Figure 3 from [4] shows what is expected from bb/cc/$\tau\tau$/gg/$\gamma\gamma$ modes, which will therefore be accessible in e+e- and should allow understanding what is going on. WW fusion in e+e- will also be measurable however, since BR(WW*)~ 1 %, a determination of the total width will be challenging.

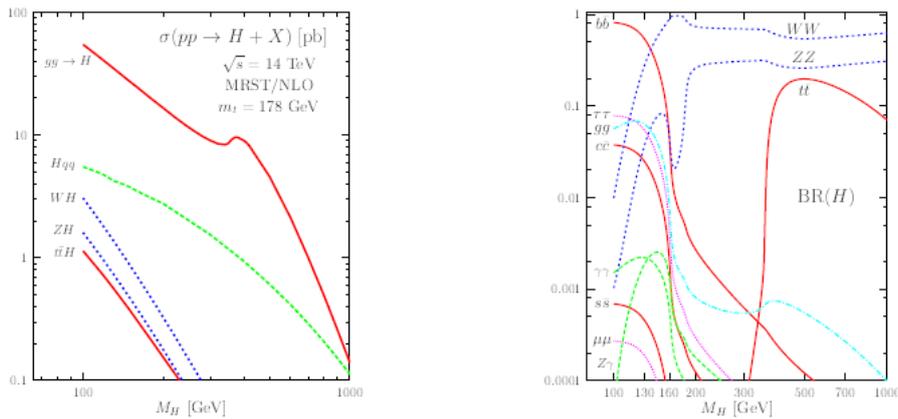

Figure 3: Cross sections and branching ratios for a SM Higgs vs. its mass, from [4]

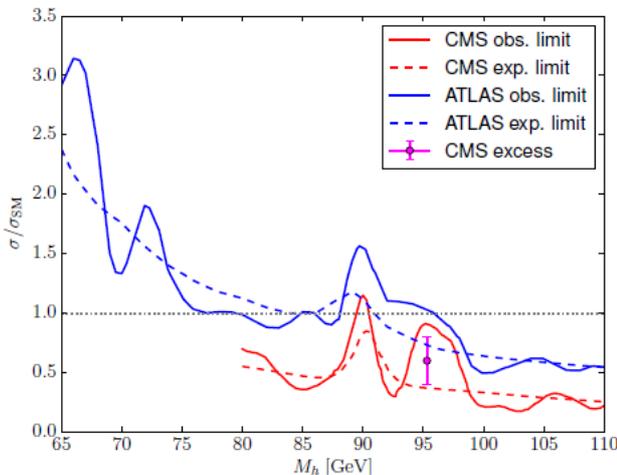

An immediate question arises: what about ATLAS? No significant evidence for this bump appears in these data. However, as pointed out in [5], the sensitivity of ATLAS [6] does not reach that of CMS, in spite of the 80 fb-1 accumulated, therefore imposing no conclusion.

One is eagerly awaiting for more data to be soon analysed. As for the interpretation, too little is known to pronounce any definite statements.

Figure 4: From [5], limits on the cross section gg->X->2$\gamma$ normalized to the SM value as a function of mh from ATLAS (blue) and CMS (red). Expected limits are in dashed lines.

Given that we cannot measure the total width, an explanation as a scalar Higgs or as a radion remain viable.

# III. X->2Z in four leptons at ATLAS and CMS

[7] took the initiative to combine available data of the two collaborations on pp->ZZ->4charged leptons and claims an evidence of 5 sd for a wide resonance around 700 GeV.

Caution is of course mandatory and one should wait for more data, as claimed by the author himself, but, in this period of 'meagre cows' and dramatic strategic decisions, one cannot disdain this emerging hope.

If real, what could it be?

Had the h(125) not been discovered, it would a priori seem an ideal SM Higgs candidate as claimed by [7], but we will see that there are significant differences.

The most obvious question is whether this new object couples to W pairs, as does h. From the observed ZZ excess, one can predict the rate of WW for a standard Higgs, with XWW/XZZ~2, and we will see that there is no compatible excess seen in ATLAS and CMS.

Does it couple to top pairs? From CMS data discussed below, this cannot be excluded given the expected interference with the gg->tt background. Since ATLAS [8] gives a hint of excess in ZZ for ggF, this suggests that the alleged resonance couples to top pairs, like for an ordinary Higgs state.

## III.1 Observations

Ref [7] interprets available data in CMS [8] and ATLAS [9] as showing two compatible excesses wider than the mass resolution. By combining these data, it claims a 5 sd effect.

This is shown in figure 5 where the SM backgrounds are below observation in the two experiments.

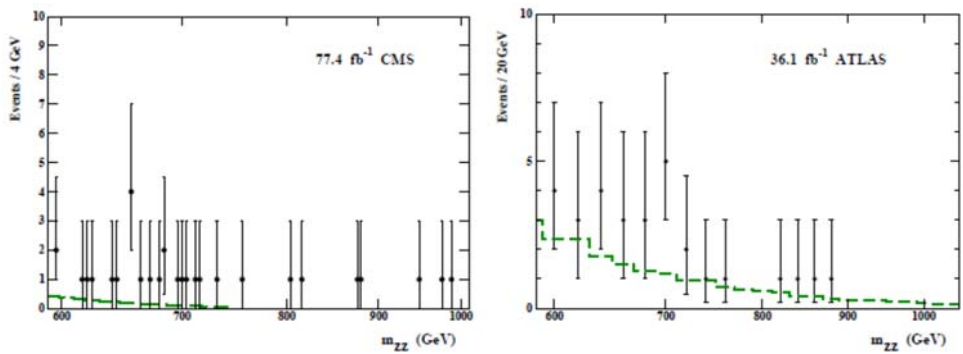

*Figure 5: Number of ZZ events into 4 leptons from CMS and ATLAS. In green, the expected background. Note the different binning in mass between CMS and ATLAS.*

Figure 6 is obtained by adding the data of figure 5. A simple adjustment, assuming a single resonance, gives:

$$M_X=660\pm10 \text{ GeV and } \Gamma_X=95\pm5 \text{ GeV}$$

Note that the mass value and the width seem incompatible with the two-photons narrow width indication observed with the first LHC data. This indication was interpreted as a narrow resonance centred at 750 GeV (see [10] and references therein).

From the combination shown in figure 6, one expects 14.2 events between 630 and 730 GeV and observes 42 events.

Open questions are:

- Do we observe the signal in tt ?
- In WW ?
- How is it shared between ggF and VBF ?

Before moving to an interpretation, I will go through some accompanying evidences and possible contradictions.

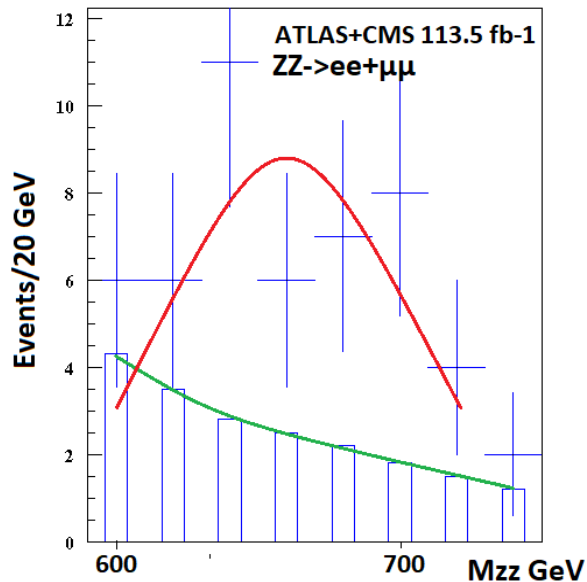

*Figure 6: Combination of CMS and ATLAS data for the ZZ measurement. In green the expected background. The red curve is a BW adjustment of the excess above background.*

## III.2 ggF and VBF from ATLAS

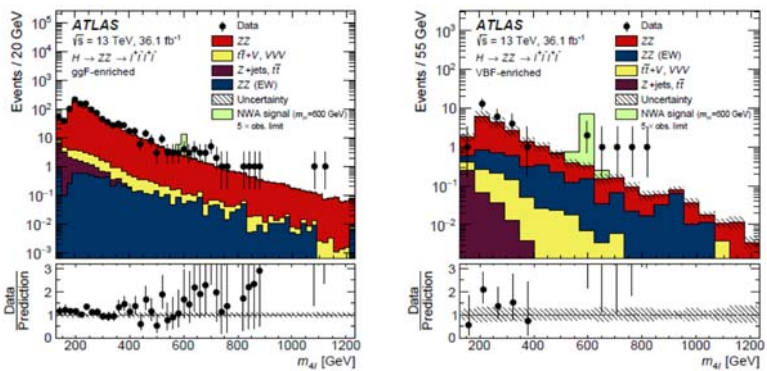

ATLAS data [9] allow interpreting the composition of the signal. The excess at 660 GeV is present in ggF and VBF.

The VBF enriched component corresponds to about 30% of the candidates, with a large uncertainty. It stands nicely above the background. At 600 GeV, the acceptance for a signal is about 34% (26% for the VBF enriched category).

*Figure 7: ggF and VBF-enriched ZZ to 4 leptons versus the reconstructed*

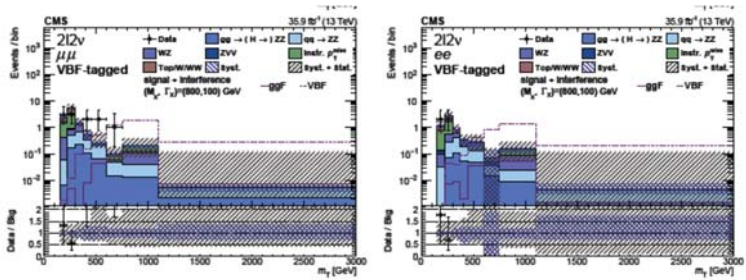

Assuming the same acceptance for CMS, one gets a ZZ cross section of 28/.005*0.34*113.5=145fb. Correcting for the BW tails (see section IV) this gives ~160 fb. This is to be compared to a SM ZZ cross section for a Higgs for a mass of 700 GeV of ~400 fb.

Figure 8: CMS searches for ZZ in ℓ+ℓ-νν, with VBF tagging

This channel has also been searched by CMS in ℓ+ℓ-νν [11], where 4 events have been observed at the relevant mass in the μμ channel and zero in ee, in the VBF enriched channel, where s/b is optimal.

### III.3 X(660)->WW

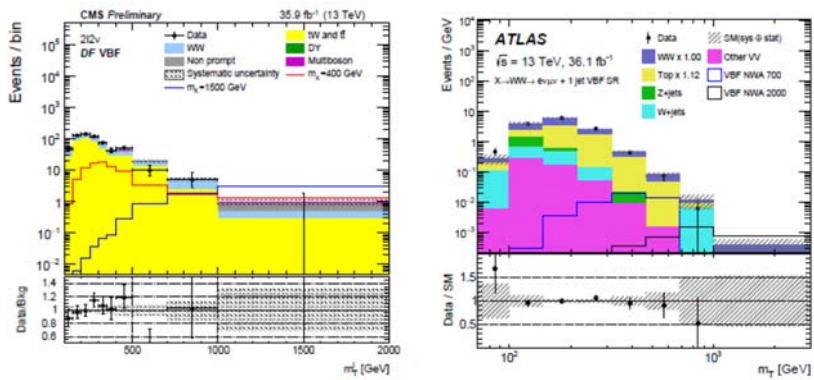

This mode has a potentially higher branching ratio for leptonic events of opposite nature, which gives a ratio $2(1/9)^2/0.005=5$ in favour of WW and, if one assumes XWW/XZZ=2, a ratio 10.

Figure 9: VBF-enriched WW events into 2l2ν versus the reconstructed mass from CMS and ATLAS data with 36 fb-1. Note the large mass resolution expected for a resonance.

The mass resolution is worse than for the four leptons case, which however, should not compensate this factor 10. Taking the VBF selection from CMS [12] and ATLAS [13], one does not observe any excess around 700 GeV. The level of background is 25 events between 500 and 1000 GeV for CMS while one would naively expect an excess of ~30 events for 36 fb-1 for a similar reconstruction efficiency. This seems excluded.

It is conceivable that only ZZ is coupled to X(660) and not WW, as will be discussed in section IV. This clearly fully distinguish this particle from an ordinary Higgs boson.

### III.4 Searches in γγ and Zγ

From CMS [14], present limits are recalled in figure 10. It seems possible to exclude the presence of a wide resonance in two-photons in the vicinity of 660 GeV, with a cross section above 3 fb, incompatible with the naïve prediction from section IV.
ATLAS [15] has a similar sensitivity, figure 11, but only provides narrow width limits.

For what concerns the Z-photon channel [16], the exclusion is less constraining, ~30 fb, but also incompatible with the naïve prediction.

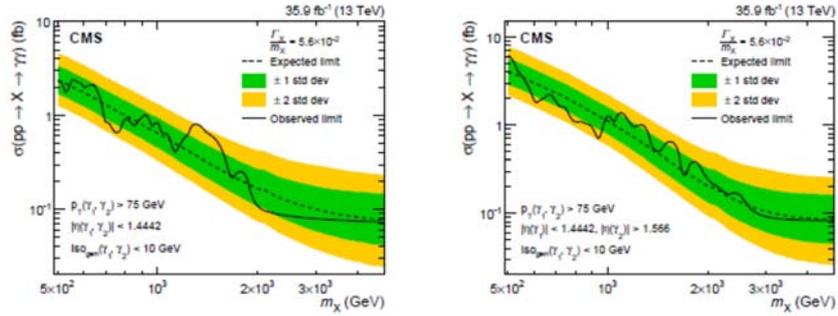

*Figure 10: From CMS, expected and observed 95% CL upper limits on pp->γγ for two categories of events.*

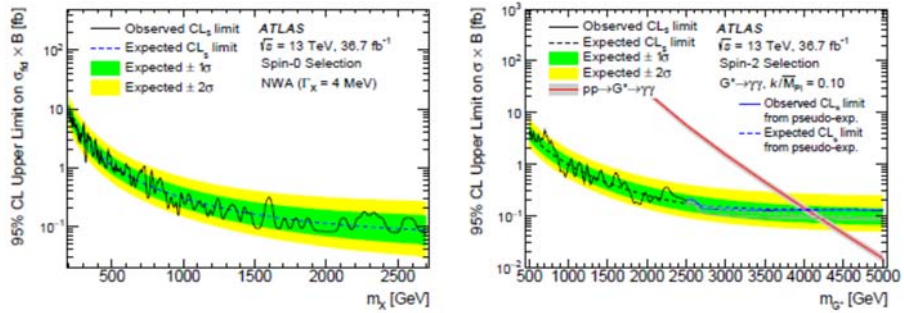

*Figure 11: From ATLAS, expected and observed 95% CL upper limits on pp->γγ for spin 0 (left) and spin 2 (right) narrow resonances.*

## III.5 Search in tt and ττ

This type of search for heavy scalars in top pairs is hampered by two facts:

- There is a huge background originating from gluon-gluon annihilation into top pairs, which, for masses ~700 GeV, overwhelms the expected signal.
- Both the signal and the background originate from gluon-gluon annihilation, which creates important interference effects that may become negative as shown by figure 12 extracted from [17] and [18].

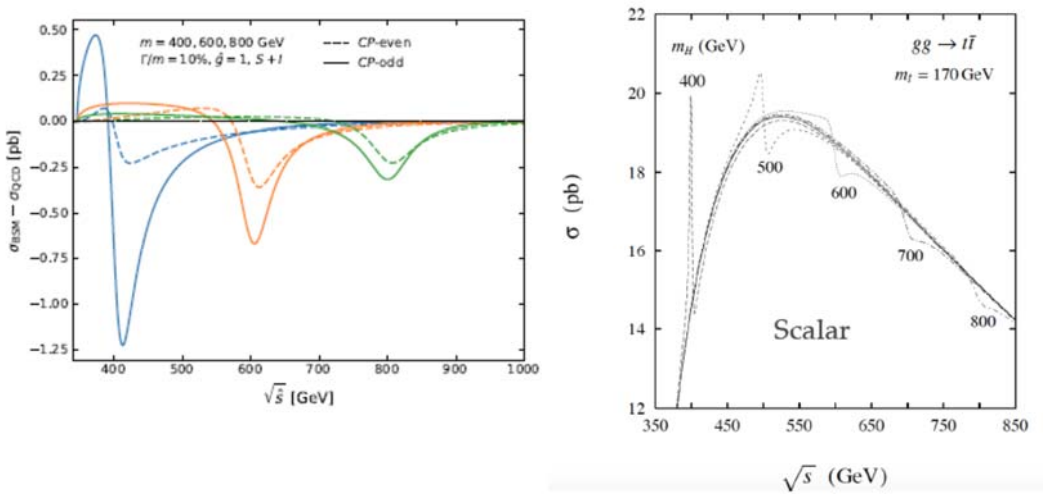

*Figure 12: Predicted interference [18] between CP-even and CP-odd putative signals and the SM model process gg->tt versus mass for 3 Higgs masses: 400, 600 and 800 GeV. The right-hand side figure shows the interference pattern expected for various scalar masses. This figure, from [17,] suggests that, at threshold, a scalar resonance would still stick out.*

At the tree level, this process depends on three interfering diagrams:

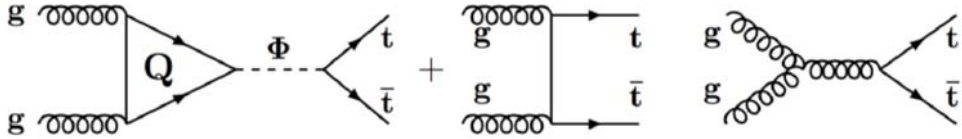

The resulting effect, shown in figure 12, shows a dominant negative interference effect. Moreover, this pattern is angular dependent since the scattering diagram has an angular variation distinct from the annihilation one, as rightly noted by CMS, which has produced mass spectra for various angular regions.

In figure 13, CMS data [19] indicates that **3.5 sd local excess at ~400 GeV**, for a total relative width of 4%, with 1.9 sd global significance (look elsewhere effect). Around 700 GeV, there is an exclusion better than expected, suggesting a deficit of events which could be due to an interference effect.

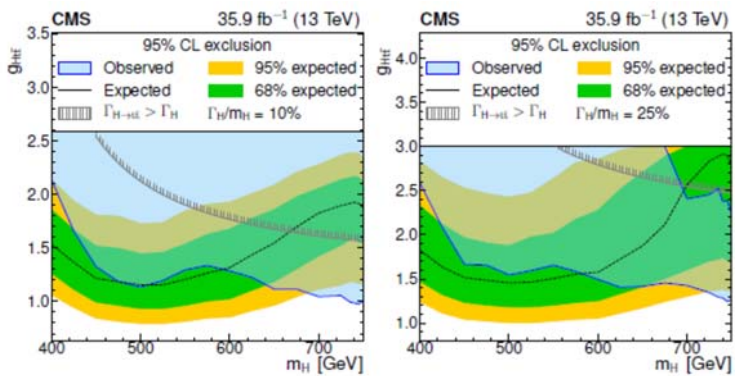

With this small fraction of the luminosity collected by CMS, one cannot draw any significant conclusion.

*Figure 13: Limits set by CMS on pp->tt, for two hypotheses on the width.*

Recently, ATLAS [26] has also produced an update on searches for H/A->ττ using 139 fb-1. Quoting the conclusions of this paper, one observes a 2.2σ excess at 400 GeV in ggF and a 2.7σ excess in H/A->ττ+bb at the same mass.

## III.6 Searches for A->Zh

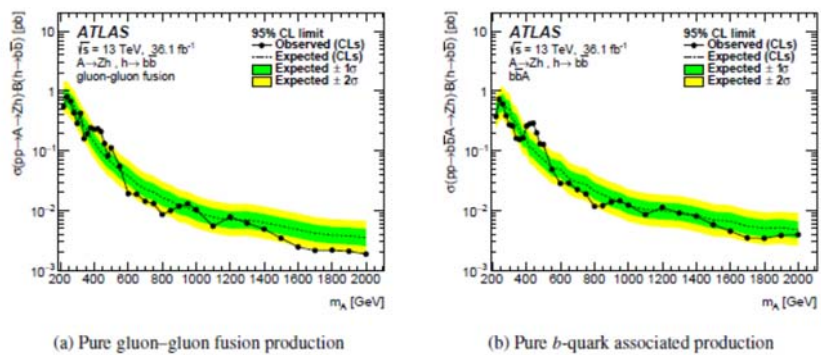

*Figure 14: Limit set by ATLAS on pp->ZA for the gluon-gluon process (a) and for the bbA associated production (b).*

From searches into top pairs, it is not easy to ascertain whether this resonance is CP-even or CP-odd. In Zh, the signature is unambiguously a CP-odd, called A. ATLAS [20] has searched for such a channel in two production mechanisms, the usual gluon fusion process, ggF, and the b-quark associated process, bbA, where A is radiated from a b quark. h(125) is detected in bb and Z into charged leptons and neutrinos.

They have and have observed a 3.6 sd local excess (2.4 sd global) at ~400 GeV in the bbA mode(figure 14 and 15). A smaller excess is observed in ggF at the same mass.

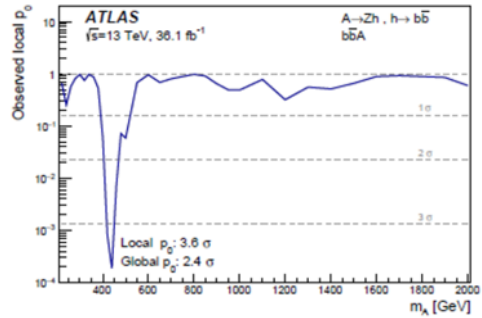

Figure 15: The observed probability (local $p_0$) that the background-only hypothesis would yield the same or more events in the combined fit for the A boson in production with associated b-quarks. For $m_A$ ~ 400 GeV, where the largest excess is observed, the local and global $p_0$ are quoted in the plot.

### III.7 Search for X(660)->h(125)h(125)

In search for a h(125)h(125) resonances, ATLAS has released, with 126 fb-1 integrated luminosity [21], the results shown on figure 16 for the VBF production. For a spin-0 broad resonance, one has a hint of an effect which could be due to X(660)->hh. One may suggest adding to the hh selection the ZZ selection in 4b. Also relevant would be a search into combinations including X(96). The mode ZA(400) is also meaningful.

The excess is about ~20 fb for VBF, to be compared to our crude estimate of 40 fb for the ZZ cross section due to VBF (see section IV).

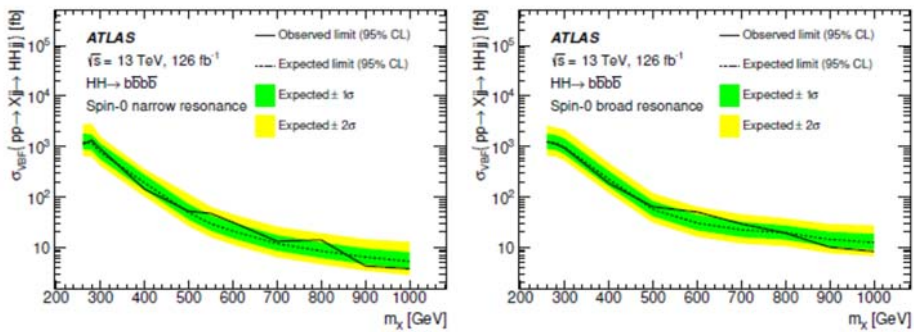

Figure 16: Observed and expected 95% CL upper limits for the resonant hh production via VBF, as a function of the mass mX. The left-hand side figure is for a narrow resonance while the right one is for a broad spin-0 resonance.

## IV. Interpretation

### IV.1 Cross section and widths of X(660)

In 4 leptons events, one has observed an excess of 42-14=28 events between 630 and 730 GeV. Since the corresponding width is 95 GeV, with an estimated mass 660 GeV, the fraction of signal included in this interval of the BW is 2[atan(30/47.5)+atan(60/47.5)]/π=0.93, hence a ZZ cross section:

$$\sigma(pp->X(660)->ZZ)=160\pm40 \text{ fb}$$

This cross section comprises two parts:
- The ggF part which is proportional to $\Gamma_{gg}BR(X \to ZZ)$
- The VBF part which is proportional to $\Gamma_{ZZ}BR(X \to ZZ)$ (where one assumes that there is no WWF)

From ATLAS data, one can guess that about 73% of the cross section is contained in the ggF part. Therefore, one takes for the ggF cross section for ZZ 0.73*160=117 fb. The VBF part is then 43 fb. This quantity is proportional to $\Gamma_{ZZ} BR(ZZ) = \Gamma^2_{ZZ}/\Gamma_{tot}$. Knowing $\Gamma_{tot}$, one extracts:

$$\Gamma_{ZZ}=27 \text{ GeV and hence BR(ZZ)}=28\%$$

These results have two consequences:

- BR(ZZ)=28% says that **a large fraction of the decay modes have not been observed**
- **$\Gamma_{ZZ}$=27 GeV** is inconsistent with an MSSM picture for which X(660) decouples from ZZ to avoids strong deviations in the h(125)ZZ coupling

From the preliminary indication shown in figure 16, one infers that the decay into **h(125)h(125)** is about ½ that in ZZ. One could also expect **contributions from X(96) pair production and from ZA(400)**, which should be searched at LHC.

Up to this point, one takes the Higgs case as a reference to identify the differences. There is however another major difference: the **absence of a WW signal**. To understand this behaviour, one can postulate that X(660) is a composite particle, a bound state $\alpha\bar{\alpha}$, where α is a colourless constituent (this aspect will become clear after the evaluation of the gluon-gluon width) with zero weak isospin. The amplitude of the process $\alpha\bar{\alpha} \to Z$ goes like $-eQ_\alpha s^2 w/swcw = -eQ_\alpha \tan w$, where $Q_\alpha$ is the charge of the constituent α. One then writes:

$$\Gamma_{ZZ}=(e^2Q^2_\alpha \tan^2 w)^2 M^3/\Lambda^2 \quad \Gamma_{\gamma\gamma}=(e^2Q_\alpha^2)^2 M^3/\Lambda^2 \quad \Gamma_{Z\gamma}=2(e^2Q^2_\alpha \tan w)^2 M^3/\Lambda^2$$

where M is the mass of this resonance and $\Lambda$ characterizes the scale of compositeness. One has:

$$\gamma\gamma/ZZ=(1/\tan^2 w)^2=11 \quad Z\gamma/ZZ=2/\tan^2 w=6.7$$

These relations will lead to a large BR into two-photons and Z-photon, which, again, is **distinct from the SM Higgs behaviour** where top and W loops contribute to a very small two-photons BR. One immediately sees that a **capital issue** will be to establish/exclude this resonance into two-photons. I recall that these large BR are incompatible with the data.

This disagreement suggest that there could be a different explanation to the origin of this bump measured in ZZ. There are many avenues in what is called "compositeness". Some, [22], [23] and [24], consider that composite scalars could be bound states of SM vector bosons or Higgs bosons. If, for some reason, X(660) would dominantly be a hh bound state, one could naïvely expect dominant decays into hh and $Z_L Z_L$, recalling that the equivalent theorem says that $Z_L \sim h$.

Reference [24] is the most predictive in treating the possibility of a '**Higgsinium**', in the framework of a two-doublet model. It sets an upper bound mA<617 GeV, by perturbativity arguments. The upper bound on H is quite similar if tanβ is large but one can guess that 660 GeV is acceptable for values like tanβ ~ 2, yet not excluded by the data. This would not be the case for tanβ=5 where mH<500 GeV. But, again, this conceptually attractive idea should not be taken literally since, within MSSM, it predicts a decoupling of the heavy Higgs to ZZ, in order to avoid perturbing the hZZ coupling. This obviously contradict the large ZZ width indicated by the data.

The next issue is to determine the ratio tt/ZZ, the top pair contribution (we will assume that other fermions contribute negligibly as for the SM Higgs case). One has tt/ZZ=$k_1 Y_X^2/\Gamma_{ZZ}$, where $k_1$ is a known coefficient and where $Y_X$ stands for the Yukawa coupling of X(660) to top pairs.

As for the SM Higgs, let us assume that the coupling to a gluon pair goes through a loop of top quarks. The ggF cross section can be written as $k_2 Y_X^2$=117 fb/BR(ZZ)=420 fb, where $k_2$~800fb. 0ne obtains **$Y_X$=0.72**, which is close to the SM case. One can also deduce the width into two gluons **$\Gamma_{gg}$ ~24 MeV**, which clearly shows that the coupling is not direct but goes through a tt loop, as for the Higgs. Therefore, the alleged **constituent of X(660) is colourless**.

| Channel | ZZ | hh | Missing | tt | gg |
|---|---|---|---|---|---|
| BR % | 28 | 14 | 49 | 9 | 0.026 |
| σ fb | 160 | 80 | 280 | 50 | 0.15 |

From this table, one easily deduce that the total production cross section for X(660) is 160/0.28=570±140fb. The missing BR could be due to A(400)Z (see section V) and/or to the bb final state. In my work [34] about the A(400) resonance, I have stated that this resonance is copiously produced in the associated process A(400)+bb, indicating that A(400) has a large Yukawa coupling to the b quark. This could also be the case for X(660).

Needless to add that this coarse evaluation allows for a large margin of uncertainty.

## IV.2 Further remarks

From a naïve constituent model, one has $\Gamma_{\gamma\gamma}$/M=45%, which would seem absurdly large according to [10]. The data invalidate this prediction and set an upper limit of order 0.1%.

If, instead, one assumes that X(660) behaves as a ZZ/hh bound state, with a new strong binding force, X(660)->hh looks the most promising mode. Figure 16 indicates that this mode could contribute like $\Gamma_{hh}/\Gamma_{ZZ}$~0.5, with large uncertainties, which is welcome to explain partly the total width. Again, a contribution from X(660)->ZA(400) and X(660)->bb are also possible.

## IV.3 Summary on X(660)

So far, one can say the following:

- There is > 4 sd excess (local) at 660 GeV in the ZZ mode in 4 charged leptons, seen by combining CMS and ATLAS data
- This resonance is wide, ~100 GeV, which excludes interpretations like a radion or a KK graviton
- The production mechanism seems to be shared between ggF and VBF, as indicated by ATLAS, with a total ZZ cross section of 160±40 fb
- There is no sign of WW final states, which should restrict possible interpretations of this resonance

- The absence of signal in two-photons and Z-photon is also a challenge for these interpretations
- There is no clear direct signal in tt, but this was anticipated due to interference with the SM background
- In the 'Higgsinium' two-doublet model, A(400) could be a CP-odd partner of X(660)
- A first hint in the hh mode for the VBF production mechanism is welcome to explain the total width of this resonance and comforts the 'Higgsinium' interpretation

From these observations, one tends to conclude that this heavy scalar is not, by far, a simple replica of the SM h(125) boson discovered at LHC.

## IV.4 A(400)

From [20], one deduces a cross section of ~330 fb for pp->(A->Zh)+bb where h is radiated from a b quark and h->bb. This excess peaks at ~440 GeV. There is also a ~2 sd excess in gg->A->Zh at about the same mass, which corresponds to a cross section ~200 fb.

In [26], the excess in $\tau\tau$ would correspond to ~30 fb while for tt the interpretation suffers from possible interference effects.

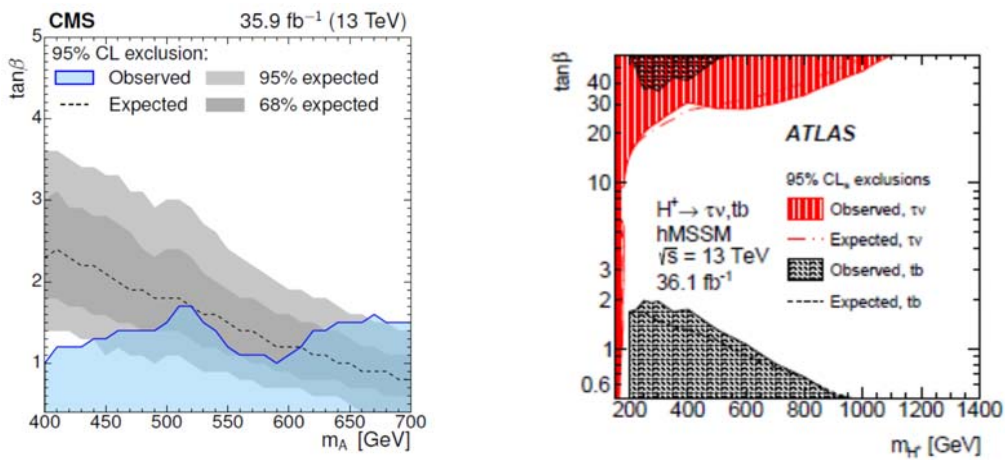

*Figure 17: On the left, excluded regions from pp->tt searches in CMS. The dashed curve shows the expected contour while the blue region is the experimental result. The discrepancy observed at 400 GeV corresponds to an excess of tt events in the CMS data. On the right, excluded regions from ATLAS obtained from pp->tb and $\tau\nu$ searches.*

There is no proof that A(400) is accompanied by a mass degenerate CP-even scalar H and by a charged H± as would happen in two-doublet model [25].

An interpretation of X(660) as a ZZ/hh/ZA bound state suggests, similarly, that A(400) could result from a Zh bound state.

How does this picture stand with the searches for A->tt and H+→tb and $\tau\nu$ from ATLAS [27]? Figure 17 shows the interpretation from [19] and [27]. One sees that the charged Higgs search leaves open a wide domain in the tanβ region, which is compatible with the excess observed in A->tt. One concludes that the solution tanβ~2 and mA~400 GeV, suggested by the A->Zh, analysis is consistent with these searches.

Recall, however, that this MSSM picture may be misleading in a composite world.

# V. An inclusive search

## V.1 Strategy

A scalar X can be observed in association with W/Z which can provide a lepton tagging:

- In genuine associated production X+W/Z
- In cascade processes like Y->X+W/Z where Y is a heavier scalar charged or neutral. If Y is neutral, then X is CP-even, while for Y charged, X can CP-even or CP-odd.
- tt+X where one can tag X by a lepton associated to top decays

For instance, one can expect X(660)->X(400)+Z, Z has a high pt of order 200 GeV, and therefore the lepton pair from Z easily passes the pt>60 GeV requirement of an ATLAS analysis [29], while X(400) decays into two jets.

If X(660) has a charged scalar partner, at a similar mass, one can have **X±(660)->X(400)+W**. This mode is more easily fulfilling the tagging requirement since the branching ratio into leptons is ~3 times larger for a W than for a Z.

This method allows to eliminate a large part of the QCD background, with the exception of top pairs.

Another advantage is that the accompanying lepton helps in passing the trigger requirement at a collider experiment.

The price to pay can be understood by looking at figure 3. Basically while associated production X+W/Z and ttH die away for masses above 200-300 GeV or so, this is not the case for the cascade process till masses reaching a TeV for gg and W/Z fusion.

## V.II Results

Reference [29] has recently explored this approach for masses above 200 GeV, using a sample with an integrated luminosity of 139 fb-1.

Results are shown in figure 18, suggesting the presence of two excesses at ~1.35 TeV and at ~400 GeV. The later already sounds familiar to our readers. These signals are best observed assuming a mass spread ~15%, which would correspond to a ~100 GeV total width for a 400 GeV resonance. This width may appear a bit large for a CP-odd decay, which does not couple to WW and ZZ. It seems to contradict the width observed by CMS using top pairs [19].

An important remark is to notice that, if these two particles, X(400) and W/Z, come from a cascade, it will be worth trying to reconstruct the mass of the original particle. This works well when the two leptons of the Z decays are detected. This may however require the **HL-LHC phase** to reach adequate sensitivity.

The two-jet signal observed at ~400 GeV using Z/W tagged events has a mass compatible with the indications discussed in the previous section, which reinforces the evidence. It could come through a cascade like X(660)->X(400)Z or, if there is a charged scalar at a similar mass, from X±->X(400)W.

**Charged scalars** can be produced in various ways. Single X±(600) could be produced by WZ fusion, by gb->b*->X±(600)+t and gg->X±(600)+tb. Typical cross sections [25] will be of order 100 fb for a mass of ~600 GeV. With X±(600)->X(400)+W, the corresponding final state will contain this W plus one or two additional W coming from the top decay. This circumstance is favourable for the lepton tagging method. Even assuming a 50% detection efficiency and a substantial decay mode X(400)+W, presumably allowed in a composite world, one cannot count on more than a few 10 fb contribution, which seems in tension with the excess indicated by figure 18, ~80±25 fb. This conclusion however depends on the unknown coupling H±tb.

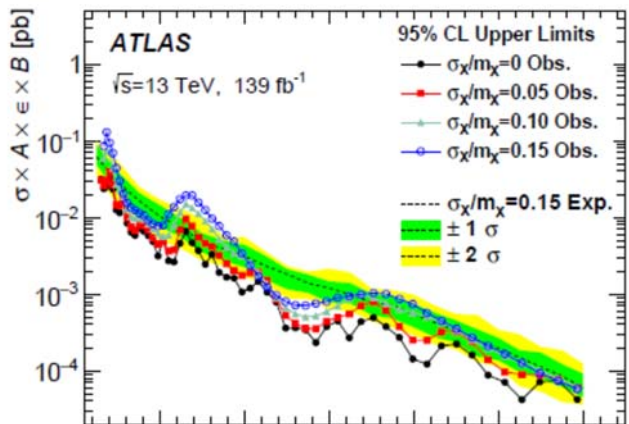

Figure 18: The 95% CL observed limits for a hypothetical particle resulting in a contribution to the 2 jet mass distribution for various Gaussian widths σx. The dashed curve corresponds to σ x/Mx=15% experimental resolution [29].

The part due to X(660)->ZA(400) would correspond to ~10 fb (see table in section IV.1).

In this type of search, interference effects between H±tb and the ttbb background may hide the signal as in the tt channel [33].

In conclusion, this indication of resonance at 400 GeV from the inclusive search of ATLAS, suggests that there could be a cascade originating from a heavy charged Higgs with a mass ~600 GeV, which decays into X(400) accompanied by a tW and tbW, providing an efficient lepton tag.

## VI.  What do we expect from HL-LHC ?

- Confirmation/rejection of above indications
- Verification of an essential feature of the ZZ leptonic modes by spin parity analysis since one expects that the Z should **be longitudinally polarized**.
- Improved sensitivities for the two-photons, Z-photon and WW modes, primordial to narrow down the interpretation of the X(660) resonance
- Search for X(660)->2h (and eventually into pairs of X(96) )
- Search for X(660)->ZA(400)
- Confirmation/rejection of the tt, $\tau\tau$ and Zh indications at ~400 GeV
- Search for a charged scalar X±(660) which cascade into X(400)+W

**Heavy vector resonances** should also be searched for, but it might take FCChh at 100 TeV to reach them.

Such resonances can also be indirectly detected, up to masses reaching 20 TeV, by precisely measuring the two-fermions cross section at ILC in the continuum and at the Z pole [28].

Aside from confirmation of some of these effects, which may already come from existing LHC data, one can hope for indirect indications coming from precision measurements of HL-LHC. If some of these scalars mix with h(125), this could already become apparent with the LHC accuracies and be conclusive with future e+e- machines.

# VII. Future e+e- machines

The four HE e+e- colliders under consideration can already settle the case for X(96) but it will take a linear collider, ILC or CLIC, to reach the heavier ones, X(660) and A(400).

## VII.1 X(96)

For what concerns X(96), as already mentioned, **e+e- is unique** to measure very precisely the ZX(96) cross section, hence the ZZ coupling, the branching ratios for bb/$\tau\tau$/cc/gg and, perhaps, $\gamma\gamma$. This will allow fully clarifying the production mechanism at LHC. It can also measure the fusion process WW->X(96), leading to a determination of the WW coupling. An access to the total width is however challenging given that BR(WW*) is at the % level.

Precision measurements are needed to understand if there is h(125)-X(96) mixing.

## VII.2 X(660)

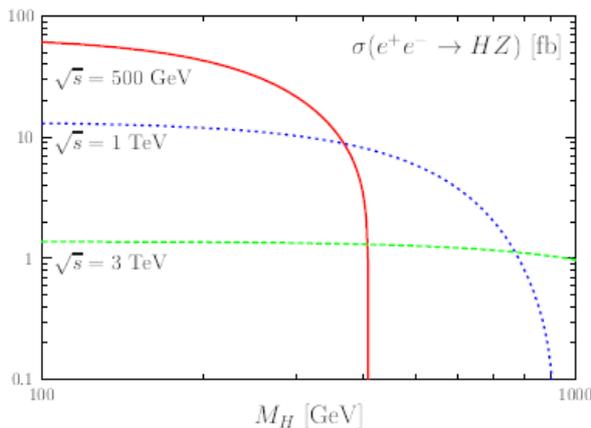

*Figure 19: From [4], cross sections for HZ versus mass, for three energies. As mentioned in the text, these cross sections have to be downscaled by two for X(660)Z.*

If true, the present observation tells that this particle couples to ZZ and therefore an e+e- machine, which can reach ~1 TeV, is adequate to produce e+e->Z*->X(660)Z. The curves from figure 18 have to be downscaled by a factor ~2 for X(660), to take into account the reduced ZZ width of X(660). From our analysis, one expects ~ 1fb at 1 TeV and half less at 3 TeV. ILC at 1 TeV could deliver 8000 fb-1, which should allow measurement accuracies at the % level.

One will very precisely measure:

- The mass
- The total width (directly since it is larger than the experimental resolution)
- The XZZ coupling
- X->tt
- X->h(125)h(125)
- X->X(96)X(96) and X(96)h(125)
- X->ZA(400), tt and $\tau\tau$
- Eventually X->2$\gamma$ X->Z$\gamma$ X->WW
- the invisible width

Recall that, contrary to h(125) where the width is not directly measurable, we do not need to access to the fusion process to determine the total width and hence derive the partial widths of the measurable processes.

The so-called NMSSM scenario [29] allows to have three neutral Higgs bosons and, assuming that H(700) is the heaviest, called h3 and X(96) the lightest called h1, one could have decays like h3->h2h2 or h3->h1h2. This scenario does not allow to have a substantial decay of h3 into ZZ.

An e+e- machine would therefore appear as the best and irreplaceable playground in these extended Higgs/composite **complex scenario**.

Note, finally, that if the composite scenario is true, in similarity to QCD with pion and rho mesons, one expects new heavy vector resonances accompanying the scalar bosons, which may require the 100 TeV pp collider to be directly observed, and an ILC to be indirectly revealed [28].

### VII.3 A(400)

The CP-odd A(400) resonance is indicated by CMS in top pairs and by ATLAS into $\tau\tau$ and Zh. The latter implies that e+e->Z*->Ah should have a finite cross section, requiring a LC operating up to 600 GeV. This resonance will be discussed in more detail in [34].

If, in spite of our reservations, one follows the analogy with MSSM where A(400) is usually accompanied by H and H± at the same mass, it would be desirable to reach 1 TeV centre of mass energy to produce e+e→HA and e+e→H+H-. Figure 20 indicate that one could expect about ~2 fb production cross section.

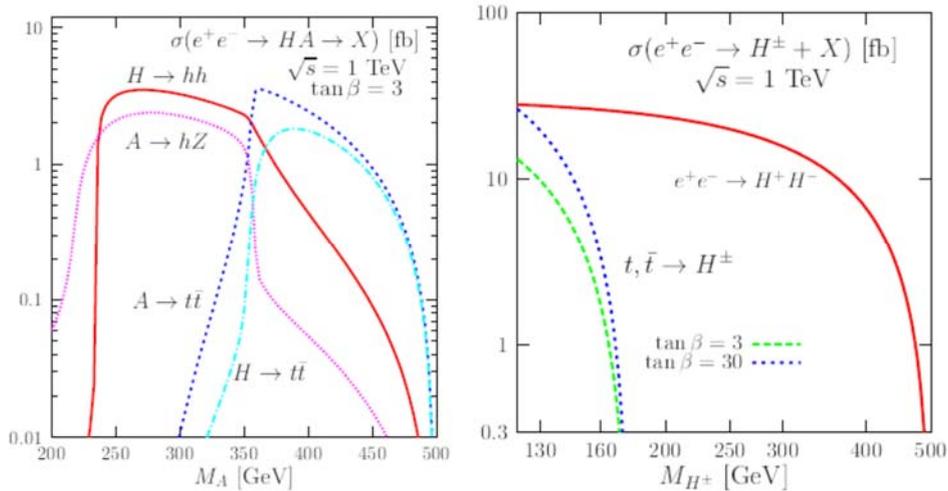

Figure 20 : From [25], on the left, cross sections times the decay BR for HA versus MA mass for tan$\beta$=3 at 1 TeV. On the right, the red curve gives the cross section for H+H-.

Figure 21, from [31], shows what can be achieved for HA and H+H-. Note that the chosen masses are below the indication at 400 GeV from ATLAS/CMS and that the integrated luminosity for H+H- is an order of magnitude (for HA two orders of magnitude !) below what is expected from ILC, while the energy is 800 GeV instead 1 TeV. Note finally that for a 400 GeV Higgs boson, the top pair channel opens up and that, unless tan$\beta$ is large, HA could decay into four top quarks, a complex channel that needs to be carefully studied.

In [34], it is however argued that with the evidence for the A(400)bb process, one expects a large coupling gAbb and hence a substantial branching ratio of this resonance into bb. This also implies that hA(400) will be easy to observe in 4b final states.

Such studies have also been performed in the CLIC collaboration [32] assuming a 4b final state from H/A, which corresponds to a large value of tan$\beta$.

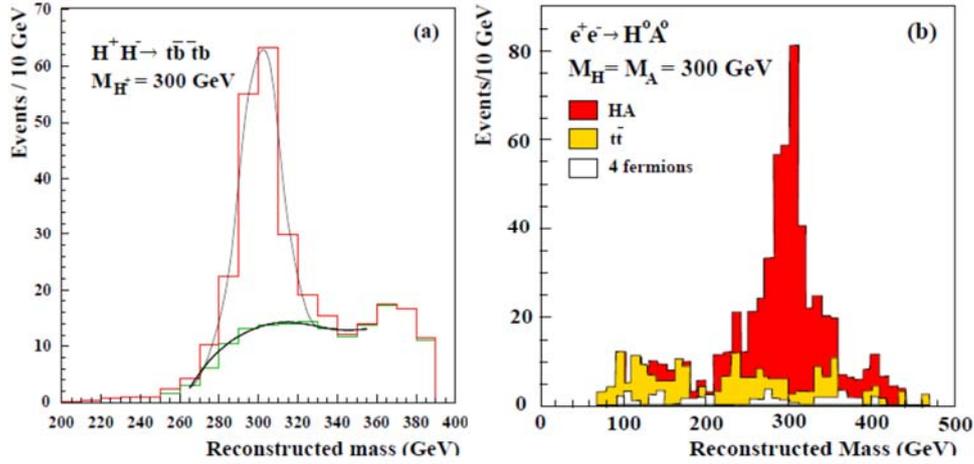

Figure 21: Left, mass reconstruction for H± decaying into tb, right, for H and A decaying into bb. These results were obtained in [31] at √s=800 GeV for 500 fb-1 for H+H- and 50 fb-1 for HA.

## VIII. Conclusion

An interesting indication for a new heavy scalar decaying into ZZ has been observed by combining ATLAS and CMS data. There is no accompanying signal into WW. If confirmed, it will provide the first evidence for BSM physics at LHC and will deserve all our attention since it may lead to a **radical change in our vision** of particle physics. Not to be forgotten is the indication in two-photons at 96 GeV from CMS and in bb at LEP2.

These indications are of course very fragile but it is comforting to know that CMS and ATLAS have already collected more data, and will collect much more with HL-LHC, which will allow to confirm/discard them. They serve as a warning for what concerns the **indispensable complementarity between pp and e+e- colliders** and the need for a LC to provide flexibility in energy to implement this complementarity.

In a naive interpretation in terms of compositeness, one expects γγ and Zγ signals, which are not observed. An alternative interpretation in terms of ZZ/hh bound states allows understanding the absence of a WW signal. **The ZZ contribution only explains partly of the X(660) width** but there is the possibility that the missing part corresponds to decays into pairs of scalars like hh – perhaps already suggested by ATLAS data – or even ZA(400).

LHC data also show indications at ~400 GeV in top pairs, in ττ, in W/Z+X, and in Zh, the latter suggesting a CP-odd particle A(400). Although each of these excesses have low statistical significance, they seem to build up into a more respectable evidence. Again, a CP-even state could contribute to these indications.

Problematic is the observation into top pairs at LHC, which could be perturbed by strong interference with the gg->tt background. This would not be the case for the VBF process.

If any of these signals is confirmed, a **rich BSM spectroscopy** of composite scalars could emerge in HEP. These scalars could then provide first answers to the fundamental problems encountered by our field and give a strong motivation to build future colliders.

**Complementarity** between pp and e+e- colliders, well proven in the case of h(125), will certainly also be essential.

**Acknowledgement and apologies**

Apologies for not citing the very numerous and inspiring theoretical papers which have provided inputs for the present document.

Acknowledgement for Philippe Zhang and his colleagues from ATLAS for providing useful information about the inclusive search with lepton tagging which is a late comer in this note in continuous progress.